\documentclass[aps,prab,twocolumn,amsmath,amssymb,longbibliography,fleqn]{revtex4-2}

\usepackage{graphicx}
\usepackage{bm}
\usepackage{mathtools}
\usepackage{siunitx}
\usepackage[hidelinks]{hyperref}

\begin{document}

\title{Rapid Mapping of Photocathode Quantum Efficiency: A Magnetized Electron Beam Imaging Approach}

\author{Yihan Liu}
\author{Lianmin Zheng}
    \email[]{zhenglm2019@mail.tsinghua.edu.cn}
\author{Yingchao Du}
    \email[]{dych@mail.tsinghua.edu.cn}
\affiliation{Department of Engineering Physics, Tsinghua University, Beijing 100084, People's Republic of China}
\affiliation{Key Laboratory of Particle and Radiation Imaging, Tsinghua University, Ministry of Education, Beijing 100084, People's Republic of China}

\begin{abstract}
Quantum efficiency (QE) is a key property of photocathodes, and its uniformity is essential for producing high-brightness electron beams. Cathode imaging provides an \textit{in-situ} and real-time approach for QE mapping, but in RF guns, a high charge per bunch is often needed to obtain a sufficient signal-to-noise ratio. Under such conditions, space charge effects can significantly degrade the imaging resolution and may even make point-to-point cathode imaging ineffective. In this paper, we propose a novel cathode imaging method based on a magnetized electron beam. Its feasibility is examined through theoretical analysis and beam dynamics simulations. The results show that the proposed method enables point-to-point cathode imaging in the ten picocoulomb charge regime. For a \SI{10}{\pico\coulomb}, \SI{3}{\pico\second} beam with a cathode magnetic field of \SI{1200}{Gauss}, simulations indicate an imaging resolution of \SI{11}{\micro\meter}, representing nearly an order-of-magnitude improvement over the non-magnetized beam method.
\end{abstract}

\maketitle

\section{Introduction}

The photocathode, mounted within the electron gun, converts photons into electrons. It generates an electron beam based on the photoelectric effect when driven by a laser. Thanks to modern advanced laser shaping techniques, the electron beam produced by the photocathode exhibits high peak current and allows precise control over its transverse and temporal distribution, making it the preferred cathode for generating high-brightness electron beams. Photocathodes have been extensively applied in numerous accelerator-based scientific facilities, such as free electron lasers \cite{emma2010first,prat2020compact,rosenzweig2020ultra,huang2021features}, Thomson scattering sources \cite{du2013generation,urakawa2011development}, and ultrafast electron diffractions \cite{filippetto2022ultrafast,nunes2020liquid}. During the operation of these scientific facilities, the uniformity of the photocathode's quantum efficiency (QE) distribution invariably deteriorates. This degradation is caused by contamination \cite{panuganti2021synthesis}, ion back bombardment \cite{grames2011charge}, and surface chemistry variations \cite{filippetto2015cesium}. Such nonuniformity in QE distribution leads to a decrease in beam brightness \cite{zhou2002experimental}. Consequently, \textit{in-situ} real-time monitoring of the QE distribution is essential to ensure the long-term stability and performance of the facilities.

QE maps can be measured with photoemission electron microscopy with resolution down to \SI{10}{\nano\meter} \cite{anders1999photoemission,polyakov2013plasmon}. However, this method requires transferring the cathode to a specialized microscopy system, which is not suitable for \textit{in-situ} QE measurements in photoinjectors. A common \textit{in-situ} approach for QE mapping is to focus the drive laser to a small spot size and raster scan it across the cathode surface \cite{schreiber2015lifetime,zheng2016development,huang2019test,panuganti2021synthesis,zhou2021commissioning,martinez2024fabrication}. The resolution depends on the laser spot size and scan step size. In practice, the scanning process takes several tens of minutes or longer. Its accuracy can be influenced by power fluctuations and laser pointing instability during this time. Digital micromirror devices \cite{riddick2013photocathode,li2017ultraviolet} have been employed in laser raster scanning to accelerate this process to several minutes. QE mapping with a \SI{47}{\micro\meter} FWHM spot size and a \SI{16}{\micro\meter} step size has been reported in Ref.~\cite{riddick2013photocathode}. Furthermore, a fast, single-shot photocathode QE mapping method has been demonstrated using a laser pattern beam with multiple beamlets \cite{wisniewski2018demonstration,zheng2020rapid}. The spatial resolution is ultimately limited by the size and separation of the individual beamlets, as well as the resolution of the imaging system.

Additionally, cathode imaging has been successfully implemented in photocathode guns to provide \textit{in-situ}, real-time QE map observations \cite{wu2017situ,liu2024adaptive}. This technique uses a solenoid as a lens to focus the electron beam and requires only single-shot electron beams to generate a QE map of the entire cathode. The imaging resolution is determined by the optical (de)magnification of the beamline and the spatial resolution of the imaging system, including the CCD sensor and scintillator screen. In DC guns, which produce continuous electron beams with relatively low beam current, space charge effects are generally negligible \cite{wu2017situ}. In contrast, radio frequency (RF) electron guns typically generate short electron bunches. Specifically, for RF guns with low repetition rates, achieving a sufficient signal‑to‑noise ratio in the imaging system when measuring QE maps using cathode imaging requires a high charge per bunch. In such cases, space charge effects significantly degrade the resolution and may even render the method ineffective.

In this paper, we propose a novel cathode imaging method based on the magnetized electron beam, aiming to provide \textit{in-situ}, single-shot photocathode QE mapping at high bunch charge. The basic concept is that the transverse coupling inherent to a magnetized beam increases the emittance, which limits the size of beam waist in the cathode imaging beamline. The enlarged waist can weaken space charge effects and suppress the distortion of the beam transverse distribution, thereby making high resolution cathode imaging feasible even in the ten picocoulomb charge regime.

This paper is organized as follows. Section~\ref{section:magnetized_beam} introduces the characteristics of the magnetized beam. The theoretical analysis of cathode imaging with the magnetized beam is presented in Sec.~\ref{section:matrix_analysis} and validated through beam dynamics simulations in Sec.~\ref{section:point_emission}. Section~\ref{section:QE_mapping} demonstrates the cathode imaging performance for high charge bunches and examines the effects of bunch charge and cathode magnetic field on the imaging quality. Section~\ref{section:summary} summarizes the current work.

\section{\label{section:magnetized_beam}Magnetized beam}

Magnetized beams, which possess significant canonical angular momentum, can be generated by applying a non-vanishing axial magnetic field $\boldsymbol{B} = B_c \hat{z}$ at the photocathode surface \cite{kim2003round, sun2004generation, kim2024four, kim2025experimental}. The magnetization is defined as the average canonical angular momentum over the beam distribution
\begin{equation}
    \mathcal{L} = \frac{L_z}{2 m_e c} = \frac{e B_c \sigma_c^2}{2 m_e c},
\end{equation}
where $L_z = \frac{e B_c}{2}\left(x^2 + y^2\right)$ represents the angular momentum imparted by the magnetic field at the cathode, $\sigma_c$ is the rms laser spot size, $m_e$ is the electron mass, and $c$ is the speed of light.

The transverse phase space vectors are defined as \(\mathbf{X}=(x,\,x')^T\),\(\mathbf{Y}=(\,y,\,y')^T\), where $x$, $y$ denote the horizontal and vertical electron positions, and $x'$, $y'$ denote the corresponding angles with respect to the longitudinal axis \(\hat{z}\). As the magnetized beam is intrinsically coupled, the four-dimensional covariance matrix can be expressed as
\begin{equation}
\Sigma = 
\begin{bmatrix}
\langle \mathbf{X} \mathbf{X}^T \rangle & \langle \mathbf{X} \mathbf{Y}^T \rangle \\
\langle \mathbf{Y} \mathbf{X}^T \rangle & \langle \mathbf{Y} \mathbf{Y}^T \rangle
\end{bmatrix}.
\end{equation}

For beams with nonzero kinetic angular momentum, the off-diagonal blocks take the form
\begin{equation}
\langle \mathbf{X} \mathbf{Y}^T \rangle = - \langle \mathbf{Y} \mathbf{X}^T \rangle = \mathcal{L} J,
\label{equ:magnetized_beam_matrix}
\end{equation}
where
\[
J \equiv 
\begin{bmatrix}
0 & 1 \\
-1 & 0
\end{bmatrix}.
\]

\section{\label{section:matrix_analysis}Matrix-based analysis of cathode imaging}

In this section, the cathode imaging beamline is simplified and examined within a transfer-matrix framework to investigate the imaging condition and the evolution of the transverse beam size along the beamline. This analysis is intended to clarify two points. First, the presence of a magnetic field at the cathode, which produces a magnetized beam, will not break the imaging condition. Second, the minimum beam size of the magnetized beam remains substantially larger than that of the non-magnetized beam, thereby mitigating the degradation of the imaging quality caused by space charge effects.

\subsection{Imaging condition}

\begin{figure}[t]
\centering

\includegraphics[width=\columnwidth]{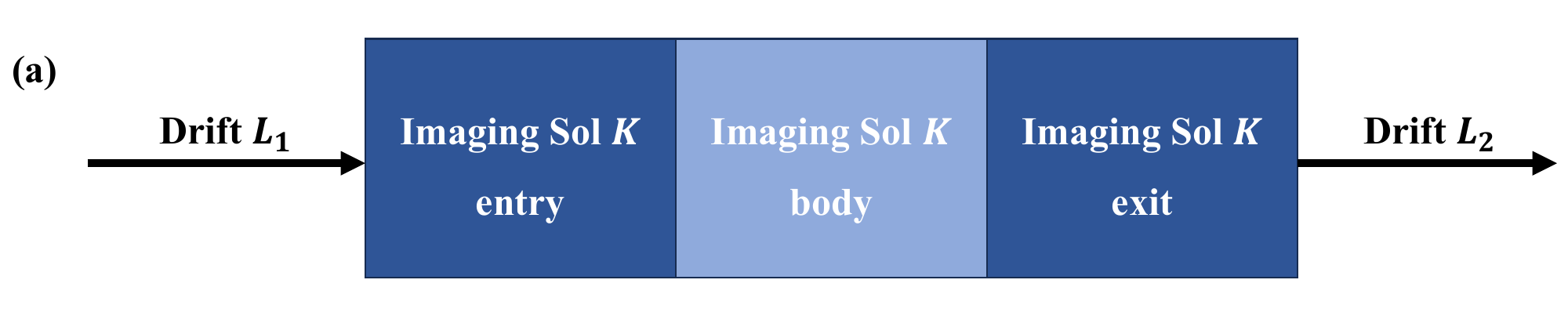}

\includegraphics[width=\columnwidth]{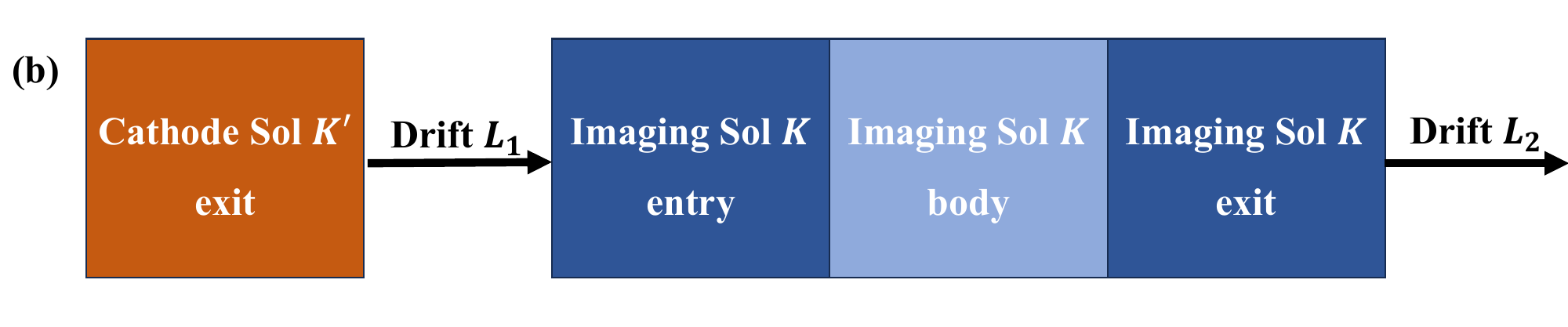}

\caption{Beamline setup used to verify the imaging condition: (a) non-magnetized beam and (b) magnetized beam.}
\label{fig:imaging-condition}
\end{figure}

The simplified beamline used to analyze the imaging condition is shown in Fig.~\ref{fig:imaging-condition}. Two configurations are considered. In the non-magnetized beam imaging beamline, as illustrated in \hyperref[fig:imaging-condition]{Fig.~\ref*{fig:imaging-condition}(a)}, the beam emitted from the cathode first propagates through a drift of length $L_1$, the imaging solenoid, and a downstream drift of length $L_2$ before reaching the observation screen. In the magnetized beam imaging beamline (see \hyperref[fig:imaging-condition]{Fig.~\ref*{fig:imaging-condition}(b)}), an additional cathode solenoid is introduced before the first drift. The strengths of the imaging solenoid and the cathode solenoid are denoted by $K$ and $K^\prime$, respectively.

To verify that the imaging condition remains valid in the presence of a cathode magnetic field, the transfer matrices of the beamline elements are constructed. The imaging solenoid is treated within the hard-edge approximation and decomposed into an entrance matrix $M_{\mathrm{entry}}$, a body matrix $M_{\mathrm{body}}$, and an exit matrix $M_{\mathrm{exit}}$. The cathode magnetic field is represented by an additional exit matrix $M_{\mathrm{exit}}^\prime$, while the drift sections are described by the drift matrix $M_{\mathrm{drift}}$.

We start from the basic transfer matrix of a hard-edge solenoid,
\begin{align}
M_{\mathrm{sol}}
&=
\begin{bmatrix}
C^2 & \tfrac{SC}{K} & SC & \tfrac{S^2}{K} \\
-KSC & C^2 & -KS^2 & SC \\
-SC & -\tfrac{S^2}{K} & C^2 & \tfrac{SC}{K} \\
KS^2 & -SC & -KSC & C^2
\end{bmatrix}
\nonumber \\
&=
M_{\mathrm{exit}} M_{\mathrm{body}} M_{\mathrm{entry}},
\label{eq:Msol}
\end{align}
\begin{align}
M_{\mathrm{exit}} &=
\begin{bmatrix}
1 & 0 & 0 & 0 \\
0 & 1 & -K & 0 \\
0 & 0 & 1 & 0 \\
K & 0 & 0 & 1
\end{bmatrix},
\label{eq:Mexit}
\\
M_{\mathrm{body}} &=
\begin{bmatrix}
1 & \tfrac{SC}{K} & 0 & \tfrac{S^2}{K} \\
0 & C^2-S^2 & 0 & 2SC \\
0 & -\tfrac{S^2}{K} & 1 & \tfrac{SC}{K} \\
0 & -2SC & 0 & C^2-S^2
\end{bmatrix},
\\
M_{\mathrm{entry}} &=
\begin{bmatrix}
1 & 0 & 0 & 0 \\
0 & 1 & K & 0 \\
0 & 0 & 1 & 0 \\
-K & 0 & 0 & 1
\end{bmatrix}.
\end{align}
where $\widehat{B}_z$, $L$, $K=e\widehat{B}_z/(2\beta\gamma m_e c)$, and $KL$ denote the axial magnetic field, effective solenoid length, solenoid strength, and Larmor angle, respectively. Here $e$ is the elementary charge, $\beta\gamma m_e c$ is the electron momentum, and $S\equiv \sin(KL)$ and $C\equiv \cos(KL)$.

The drift sections are described by
\begin{equation}
M_{\mathrm{drift}}
=
\begin{bmatrix}
1 & L_d & 0 & 0 \\
0 & 1 & 0 & 0 \\
0 & 0 & 1 & L_d \\
0 & 0 & 0 & 1
\end{bmatrix},
\label{eq:Mdrift}
\end{equation}
where $L_d$ is the drift length.

The initial beam is characterized by the sigma matrix at the cathode. For simplicity, the beam is assumed to have zero intrinsic emittance, a uniform transverse distribution, identical rms beam size in the $x$ and $y$ directions, and zero initial divergence. The initial sigma matrix can thus be written as
\begin{equation}
\Sigma_0
=
\begin{pmatrix}
\langle x_0^2 \rangle & 0 & 0 & 0 \\
0 & 0 & 0 & 0 \\
0 & 0 & \langle x_0^2 \rangle & 0 \\
0 & 0 & 0 & 0
\end{pmatrix}.
\label{eq:Sigma0}
\end{equation}

For the magnetized beam imaging beamline, the effect of the cathode magnetic field is represented by an additional exit matrix
\begin{equation}
M_{\mathrm{exit}}^\prime=
\begin{bmatrix}
1 & 0 & 0 & 0 \\
0 & 1 & -K^\prime & 0 \\
0 & 0 & 1 & 0 \\
K^\prime & 0 & 0 & 1
\end{bmatrix},
\label{eq:Mexitprime}
\end{equation}
which represents the transverse momentum kick introduced by the cathode solenoid at emission. The sigma matrix immediately after the cathode field is given by
\begin{align}
\Sigma
&=
M_{\mathrm{exit}}^\prime \Sigma_0 \left(M_{\mathrm{exit}}^\prime\right)^T
\nonumber \\
&=
\begin{pmatrix}
\langle x_0^2 \rangle & 0 & 0 & K^\prime \langle x_0^2 \rangle \\
0 & K^{\prime 2}\langle x_0^2 \rangle & -K^\prime \langle x_0^2 \rangle & 0 \\
0 & -K^\prime \langle x_0^2 \rangle & \langle x_0^2 \rangle & 0 \\
K^\prime \langle x_0^2 \rangle & 0 & 0 & K^{\prime 2}\langle x_0^2 \rangle
\end{pmatrix}.
\label{eq:Sigma_magnetized}
\end{align}

For the non-magnetized beam imaging beamline in \hyperref[fig:imaging-condition]{Fig.~\ref*{fig:imaging-condition}(a)}, the transport matrix from the cathode to the screen is
\begin{widetext}
\begingroup
\setlength{\mathindent}{0pt}

\begin{equation}
\begin{aligned}
M_{\mathrm{non-magnetized}}
&= M_{\mathrm{drift2}}\,M_{\mathrm{sol}}\,M_{\mathrm{drift1}}
\\
&=
\normalsize
\begin{bmatrix}
1 & L_2 & 0 & 0 \\
0 & 1 & 0 & 0 \\
0 & 0 & 1 & L_2 \\
0 & 0 & 0 & 1
\end{bmatrix}
\begin{bmatrix}
C^2 & \tfrac{SC}{K} & SC & \tfrac{S^2}{K} \\
-KSC & C^2 & -KS^2 & SC \\
-SC & -\tfrac{S^2}{K} & C^2 & \tfrac{SC}{K} \\
KS^2 & -SC & -KSC & C^2
\end{bmatrix}
\begin{bmatrix}
1 & L_1 & 0 & 0 \\
0 & 1 & 0 & 0 \\
0 & 0 & 1 & L_1 \\
0 & 0 & 0 & 1
\end{bmatrix}
\\
&=
\begingroup
\small
\frac{1}{2}
\begin{bmatrix}
1+c-\beta s
&
\tfrac{(1+c)(\alpha+\beta)+s(1-\alpha\beta)}{K}
&
-\beta+\beta c+s
&
\tfrac{(1-\alpha\beta)(1-c)+s(\alpha+\beta)}{K}
\\
-Ks
&
1+c-\alpha s
&
-K(1-c)
&
-\alpha +\alpha c+s
\\
\beta-\beta  c-s
&
-\tfrac{(1-\alpha\beta)(1-c)+s(\alpha+\beta)}{K}
&
1+c-\beta s
&
\tfrac{(1+c)(\alpha+\beta)+s(1-\alpha\beta)}{K}
\\
K(1-c)
&
\alpha-\alpha c-s
&
-Ks
&
1+c-\alpha s
\end{bmatrix},
\endgroup
\end{aligned}
\end{equation}
\endgroup
where $s \equiv \sin(2KL)$, $c \equiv \cos(2KL)$, $\alpha \equiv KL_1$, $\beta \equiv KL_2$.

For the magnetized beam imaging beamline in \hyperref[fig:imaging-condition]{Fig.~\ref*{fig:imaging-condition}(b)}, the cathode exit matrix should be included before the first drift, and the transport matrix becomes
\begingroup
\setlength{\mathindent}{0pt}
\begin{equation}
\begin{aligned}
M_{\mathrm{magnetized}}
&=
M_{\mathrm{drift2}}\,M_{\mathrm{sol}}\,M_{\mathrm{drift1}}\,M_{\mathrm{exit}}^\prime
\\
&=
\footnotesize
\begin{bmatrix}
1 & L_2 & 0 & 0 \\
0 & 1 & 0 & 0 \\
0 & 0 & 1 & L_2 \\
0 & 0 & 0 & 1
\end{bmatrix}
\begin{bmatrix}
C^2 & \tfrac{SC}{K} & SC & \tfrac{S^2}{K} \\
-KSC & C^2 & -KS^2 & SC \\
-SC & -\tfrac{S^2}{K} & C^2 & \tfrac{SC}{K} \\
KS^2 & -SC & -KSC & C^2
\end{bmatrix}
\begin{bmatrix}
1 & L_1 & 0 & 0 \\
0 & 1 & 0 & 0 \\
0 & 0 & 1 & L_1 \\
0 & 0 & 0 & 1
\end{bmatrix}
\begin{bmatrix}
1 & 0 & 0 & 0 \\
0 & 1 & -K' & 0 \\
0 & 0 & 1 & 0 \\
K' & 0 & 0 & 1
\end{bmatrix}
\\
&
=
M_{\mathrm{non-magnetized}}
+
\frac{K'}{2}\,
\begingroup
\small
\begin{bmatrix}
\tfrac{(1-c)(1-\alpha\beta)+s(\beta+\alpha)}{K}
&
0
&
-\tfrac{(1+c)(\alpha+\beta)+s(1-\alpha\beta)}{K}
&
0
\\
-\alpha+\alpha c+s
&
0
&
-1-c+\alpha s
&
0
\\
\tfrac{(1+c)(\alpha+\beta)+s(1-\alpha\beta)}{K}
&
0
&
\tfrac{(1-c)(1-\alpha\beta)+s(\alpha+\beta)}{K}
&
0
\\
1+c-\alpha s
&
0
&
-\alpha+\alpha c+s
&
0
\end{bmatrix}.
\endgroup
\end{aligned}
\end{equation}
\endgroup
\end{widetext}

The imaging condition requires that the transverse position coordinates on the observation screen be independent of the initial transverse momentum at the cathode. For the four-dimensional transport matrix from the cathode to the screen, this requirement can be written as
\begin{equation}
\begin{bmatrix}
x \\
x^\prime \\
y \\
y^\prime
\end{bmatrix}_{\mathrm{screen}}
=
\begin{bmatrix}
M_{11} & 0 & M_{13} & 0 \\
M_{21} & M_{22} & M_{23} & M_{24} \\
M_{31} & 0 & M_{33} & 0 \\
M_{41} & M_{42} & M_{43} & M_{44}
\end{bmatrix}
\begin{bmatrix}
x \\
x^\prime \\
y \\
y^\prime
\end{bmatrix}_{\mathrm{cathode}}.
\label{eq:imaging_condition}
\end{equation}

The imaging conditions for the two configurations are found to be identical:
\begin{equation}
\begin{split}
& L_2 = \frac{K L_1 + \tan(KL)}{K\bigl(-1 + K L_1 \tan(KL)\bigr)}, \\
& K\bigl(-1 + K L_1 \tan(KL)\bigr) \neq 0, \\
& \frac{2KL - \pi}{2\pi} \notin \mathbb{Z}.
\end{split}
\label{eq:imaging_condition_solution}
\end{equation}

Eqs.~\eqref{eq:imaging_condition_solution} shows that the imaging position $L_2$ is independent of the cathode solenoid strength $K^\prime$. Therefore, the imaging condition remains valid after the introduction of the magnetic field at the cathode.

\subsection{Transverse Beam Size Evolution}

\begin{figure}[b]
\centering
\includegraphics[width=\columnwidth]{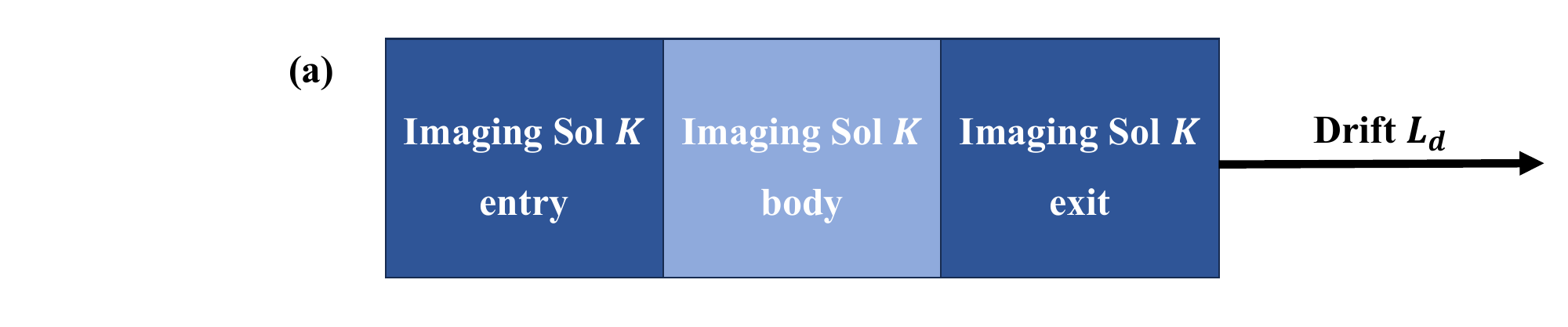}
\includegraphics[width=\columnwidth]{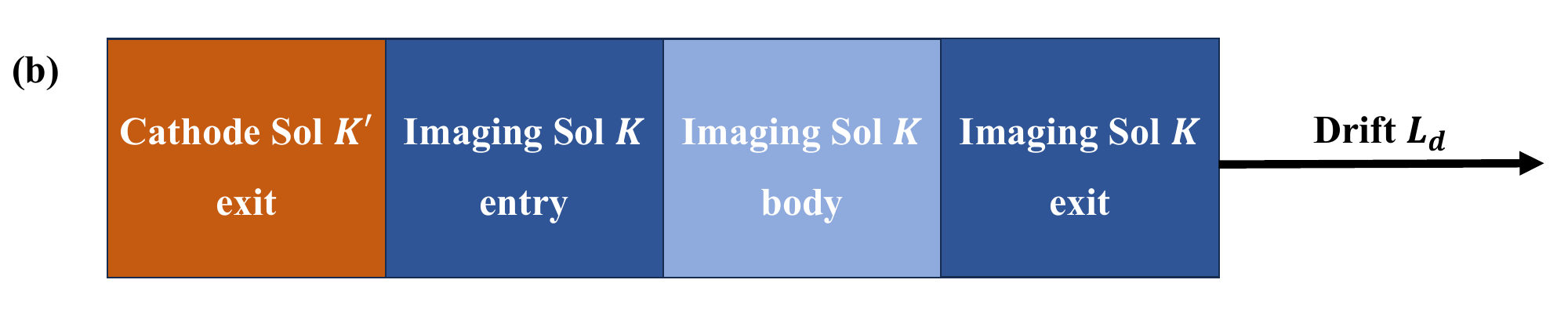}
\caption{Beamline setup used to calculate the beam envelope evolution: (a) non-magnetized beam and (b) magnetized beam.}
\label{fig:mma}
\end{figure}

Although the cathode magnetic field does not break the imaging condition, it changes the transverse beam size evolution due to the canonical angular momentum introduced at emission. Here the beam matrix formalism is employed to evaluate the rms beam size, revealing that the minimum beam size increases in the magnetized beam method.

For simplicity, the drift between the cathode and the imaging solenoid is neglected. The simplified beamlines are shown in Fig.~\ref{fig:mma}. In the non-magnetized beam method [\hyperref[fig:mma]{Fig.~\ref*{fig:mma}(a)}], the beam matrix at the end of the drift is
\begin{equation}
    \Sigma_1 = \left(M_{\mathrm{drift}} M_{\mathrm{sol}}\right) \Sigma_0 \left(M_{\mathrm{drift}} M_{\mathrm{sol}}\right)^T,
    \label{equ:无阴极磁场漂移节出口处的束流矩阵}
\end{equation}
and the corresponding squared rms beam size $\sigma_1^2$ varies with the drift length $L_d$,
\begin{align}
    \sigma_1^2 & = \Sigma_1\left(1,1\right) \nonumber \\
    & = \left\langle x_0^2 \right\rangle \left[ \left(C^2 - K L_d S C \right)^2 + \left( C S - K L_d S^2 \right)^2 \right] \nonumber \\
    & = \left\langle x_0^2 \right\rangle \left( S K L_d - C \right)^2.
    \label{equ:无阴极磁场漂移节出口处束团均方根尺寸的平方}
\end{align}

Eq.~\eqref{equ:无阴极磁场漂移节出口处束团均方根尺寸的平方} indicates that if the space charge effects are neglected, all electrons will converge at the focal point
\begin{equation}
    L_d = \frac{\cot(KL)}{K},
\end{equation}
where the beam size vanishes.

In the magnetized beam method [\hyperref[fig:mma]{Fig.~\ref*{fig:mma}(b)}], the cathode field is represented by the additional exit matrix before the imaging solenoid. The beam matrix at the end of the drift is then
\begin{equation}
    \Sigma_2
    = \left(M_{\mathrm{drift}} M_{\mathrm{sol}} M_{\mathrm{exit}}^{\prime}\right)
    \Sigma_0
    \left(M_{\mathrm{drift}} M_{\mathrm{sol}} M_{\mathrm{exit}}^{\prime}\right)^T .
    \label{equ:磁化束漂移节出口处的束流矩阵}
\end{equation}

Accordingly, the squared rms beam size becomes
\begin{align}
    \sigma_2^2 & = \Sigma_2(1,1) \nonumber \\
    & = \left\langle x_0^2 \right\rangle 
    \begin{bmatrix}
    \biggl[ S C - K L_d S^2 - K^\prime \left( L_d C^2 + \dfrac{S C}{K} \right) \biggr]^2 + \\[0.5ex]
    \biggl[ C^2 - K L_d S C + K^\prime \left( L_d S C + \dfrac{S^2}{K} \right) \biggr]^2
    \end{bmatrix}.
    \label{equ:磁化束均方根尺寸的平方}
\end{align}

For fixed $K^\prime$, $K$, and $L$, the minimum beam size remains finite,
\begin{equation}
    \sigma_{2,\min}^2
    = \frac{K^{\prime 2}\left\langle x_0^2 \right\rangle}
    {K^{\prime 2}C^2+K^2S^2},
    \label{equ:磁化束最小均方根尺寸平方}
\end{equation}
which is reached at
\begin{equation}
    L_d =
    \frac{\left(K^2-K^{\prime 2}\right)SC}
    {K\left(K^{\prime 2}C^2+K^2S^2\right)} .
    \label{equ:磁化束最小束斑位置}
\end{equation}

Numerical evaluation is performed using Mathematica \cite{Mathematica}, with the imaging solenoid assumed to have a Larmor angle of $KL=\SI{0.82}{\radian}$ and an effective length of \SI{0.3}{\meter}. For a nonzero cathode field, the canonical transverse momentum imparted at emission partially cancels the momentum change at the entrance of the imaging solenoid. Cathode solenoid strengths of $K^\prime/K=1/5$ and $K^\prime/K=1/10$ are therefore considered. The initial beam is assumed to be transversely uniform with $\sqrt{\langle x_0^2\rangle}=\SI{1}{\milli\meter}$ and zero intrinsic emittance.

\begin{figure}[tb]
    \centering
    \includegraphics[width=\columnwidth]{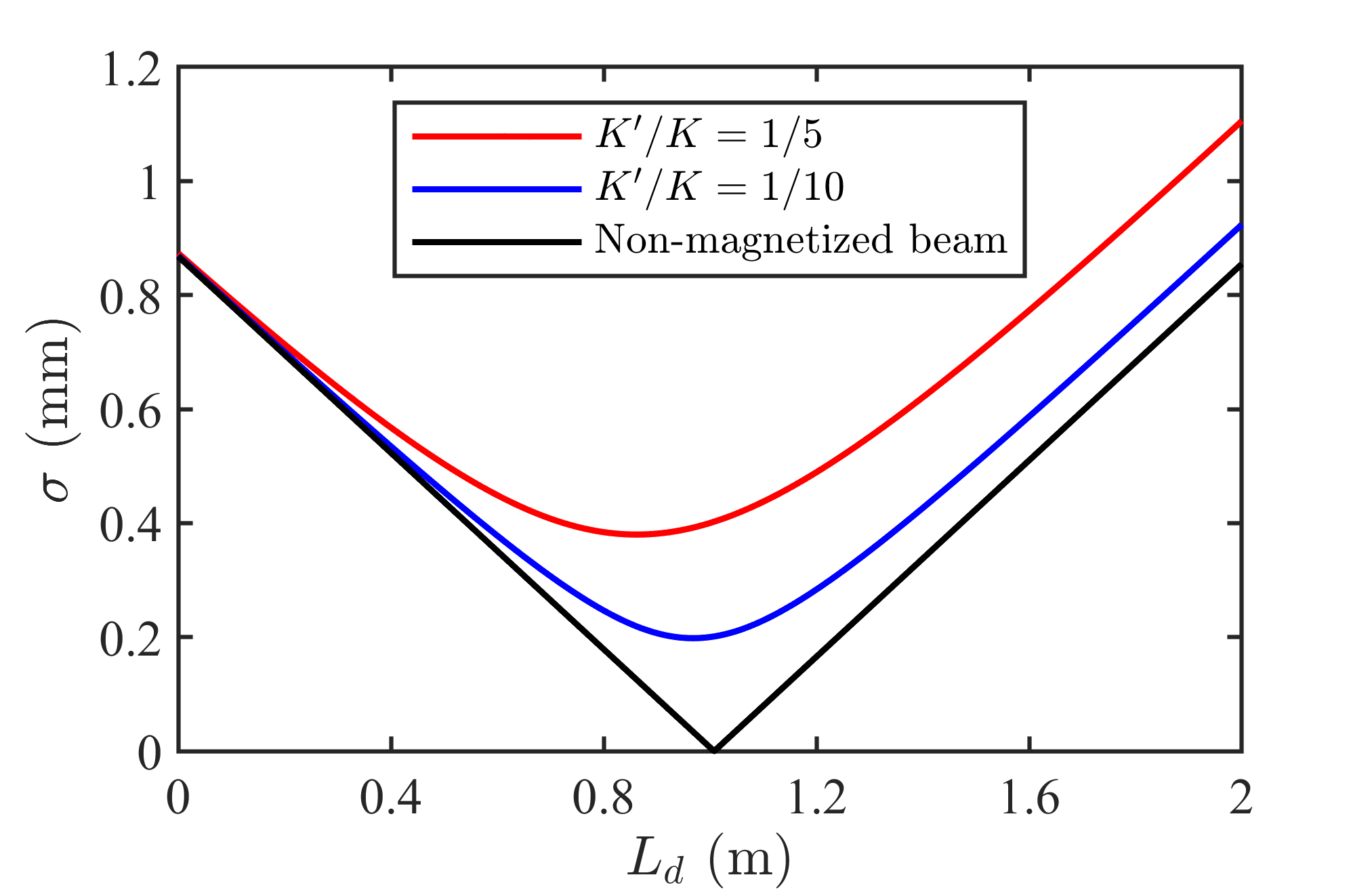}
    \caption{RMS beam size $\sigma$ as a function of drift length $L_d$.}
    \label{fig:束团横向尺寸与漂移节长度}
\end{figure}

The calculated rms beam sizes are shown in Fig.~\ref{fig:束团横向尺寸与漂移节长度}. In the non-magnetized beam method, the beam can be focused to a vanishingly small waist in the absence of space charge. In contrast, in the magnetized beam method, the minimum beam size remains finite because the canonical angular momentum increases the beam transverse emittance. This increase in beam waist size is beneficial for high charge cathode imaging: it suppresses the space charge enhancement near the focus and helps preserve point-to-point imaging.

\section{\label{section:point_emission}Beam dynamics simulations of point emission}

In Sec.~\ref{section:matrix_analysis}, the imaging condition and beam size evolution are analyzed within an ideal linear imaging system without space charge effects. In practical cathode imaging, however, the space charge effects play an important role, particularly for high charge bunches. 

In this section, beam dynamics simulations are performed with ASTRA \cite{flottmann2011astra} for both the non-magnetized beam method and the magnetized beam method. A space charge force environment containing a localized point emission is constructed to analyze the dynamics of the point-emission particles and to quantify the imaging resolution. The analysis is carried out from four perspectives: beam envelope evolution, point-emission resolution, and the trajectories and transverse momentum evolution of the point-emission macroparticles.

The simulations are carried out using the Tsinghua very-high-frequency (VHF) electron gun injector \cite{chen2021analysis,chen2022beam,zheng2023design}, which serves as a representative example for illustrating the advantages of magnetized beam cathode imaging. It should be noted that the proposed method is applicable to photoinjectors in general, especially those in which a cathode magnetic field can be generated readily without altering the beamline configuration, such as the Photo Injector Test Facility (PITZ) at Deutsches Elektronen-Synchrotron DESY \cite{huang2019test} and the Argonne Wakefield Accelerator (AWA) injector \cite{conde2017research}.

\begin{figure}[t]
\centering
\includegraphics[width=\columnwidth]{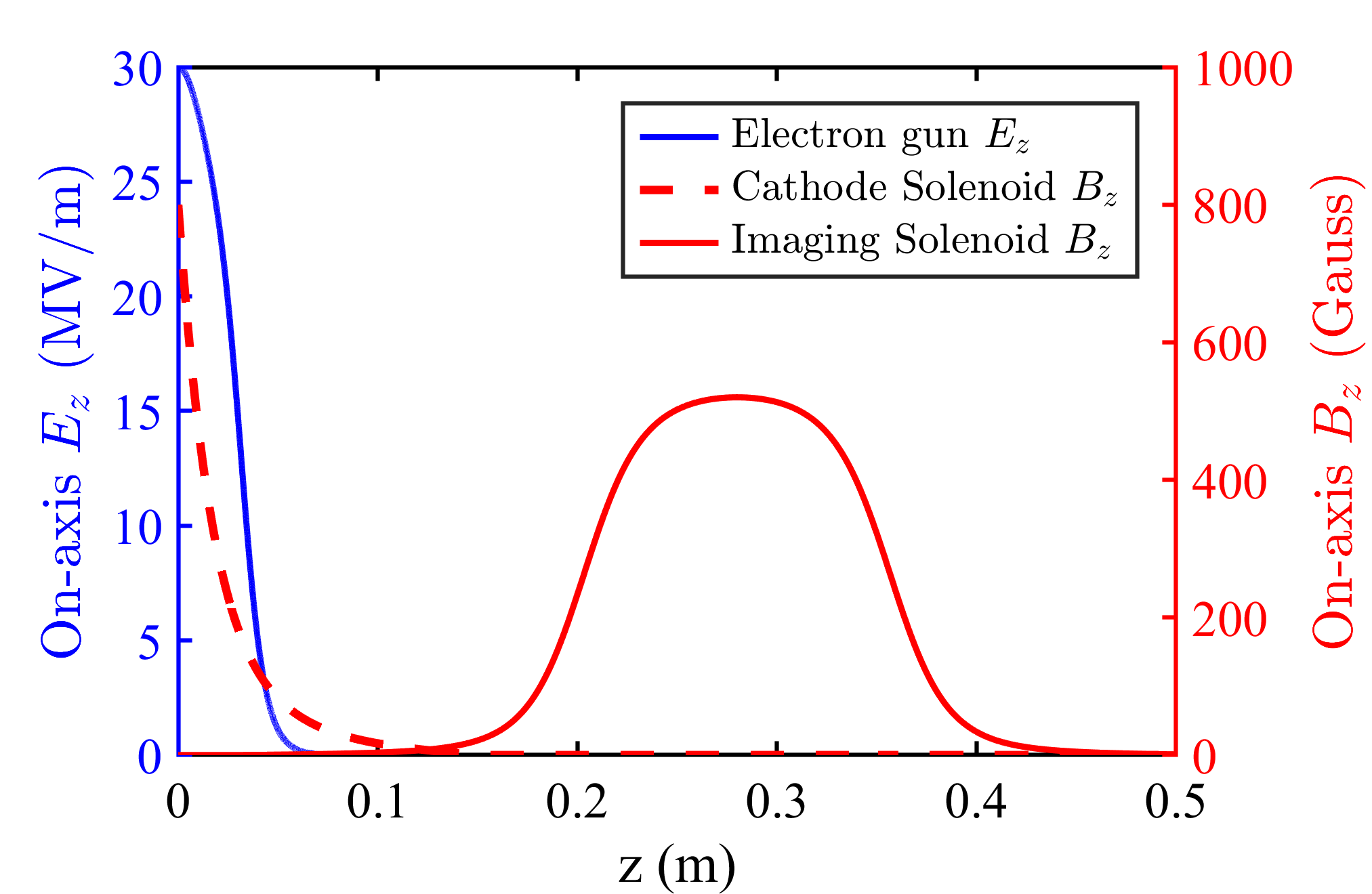}
\caption{
On-axis field profiles used in the simulations:
electron gun $E_z(z)$ field (blue solid), cathode solenoid $B_z(z)$ field (red dashed), and imaging solenoid $B_z(z)$ field (red solid).
}
\label{fig:VHF_EfieldBfield}
\end{figure}

In the simulations, the electron beam at the VHF gun exit reaches an energy of \SI{0.87}{\mega\electronvolt} and is subsequently focused by the imaging solenoid. The solenoid center, which provides the cathode magnetic field, is positioned \SI{0.10}{\meter} upstream of the cathode surface, while the imaging solenoid center is located at \SI{0.28}{\meter} along the beamline. The axial electric and magnetic field profiles are shown in Fig.\ref{fig:VHF_EfieldBfield}.

\subsection{\label{subsection:point_resolution_initial_dist}Initial beam distribution}

A transversely uniform bunch is adopted to establish an approximately uniform space charge force environment. The beam has a transversely uniform distribution and a longitudinal Gaussian profile, with isotropic emission from the cathode. Two bunch settings are considered to represent distinct space charge regimes:
\begin{align}
    \text{(i)}\quad
    & Q=\SI{10}{\femto\coulomb},
    \qquad
    \sigma_t=\SI{100}{\femto\second},
    \nonumber\\
    \text{(ii)}\quad
    & Q=\SI{10}{\pico\coulomb},
    \qquad
    \sigma_t=\SI{3}{\pico\second}.
    \nonumber
\end{align}

In both cases, the rms beam size on the cathode $\sigma_{\mathrm{cathode}}$ is set to \SI{1}{\milli\meter}, and the number of macroparticles is fixed at $62{,}500$.

To represent the localized point emission, a group of 10 macroparticles is assigned the same emission position, $(x_0,y_0)=(\SI{0.5}{\milli\meter},\,\SI{0.5}{\milli\meter})$, while their emission times and initial momentum distribution remain unchanged.

\subsection{\label{subsection:point_resolution}Beam envelope and point-emission resolution}

\begin{figure*}[t]
\centering

\includegraphics[width=0.48\textwidth]{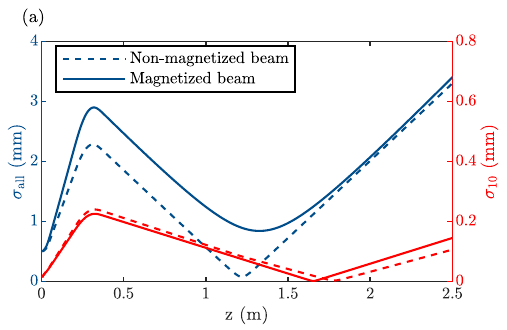}
\hfill
\includegraphics[width=0.48\textwidth]{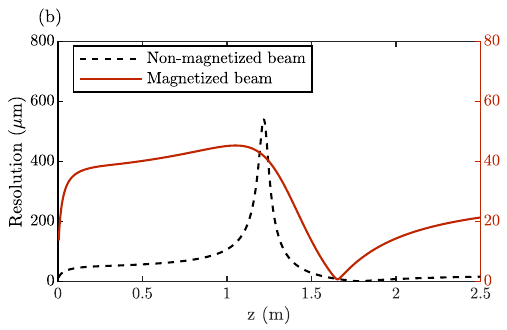}

\includegraphics[width=0.48\textwidth]{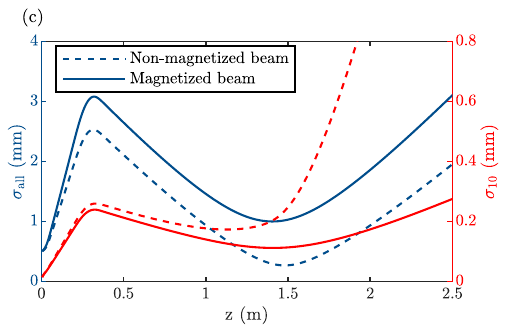}
\hfill
\includegraphics[width=0.48\textwidth]{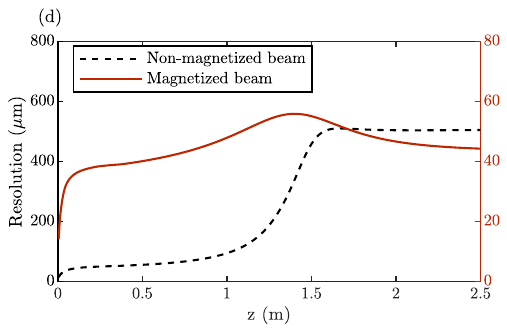}

\caption{(a) Beam envelope evolution and (b) cathode-plane resolution for the \SI{10}{\femto\coulomb} beam. 
(c) Beam envelope evolution and (d) cathode-plane resolution for the \SI{10}{\pico\coulomb} beam.}
\label{fig:point-resolution-10fC-10pC}
\end{figure*}

To characterize the dynamics of the imaging process, the positions and momenta of all macroparticles are recorded every \SI{5}{\milli\meter} along the beamline. At each sampling plane, the transverse beam distribution is characterized by two quantities:
{\renewcommand\labelenumi{(\roman{enumi})}%
\begin{enumerate}
    \item the rms spot size of all macroparticles, $\sigma_{\mathrm{all}}$,
    \item the rms size of the point-emission microbunch formed by 10 macroparticles, $\sigma_{10}$.
\end{enumerate}}

In an ideal imaging system, space charge effects are neglected and the solenoid is treated as a thin lens. Under these assumptions, imaging is achieved when the point-emission microbunch satisfies $\sigma_{10} \rightarrow 0$. In practical operation, however, non-ideal effects, such as space charge effects and higher-order multipole components in the solenoid field, cannot be ignored. Accordingly, the rms size of the 10 macroparticles point-emission microbunch at a given sampling plane is used here as a proxy for the local imaging resolution on the screen.

As the beam is focused by the solenoid and subsequently expands in the downstream drift, the magnification of the imaging system and the corresponding cathode-plane resolution are defined as
\begin{align}
  \mathrm{mag} & = \frac{\sigma_{\mathrm{all}}}{\sigma_{\mathrm{cathode}}}, \\
  \mathrm{res} & = \sigma_{10} \times \frac{\sigma_{\mathrm{cathode}}}{\sigma_{\mathrm{all}}}.
\end{align}

Fig.~\ref{fig:point-resolution-10fC-10pC} shows the evolutions of $\sigma_{\mathrm{all}}$ and $\sigma_{10}$ for the non-magnetized and magnetized beams in the two representative space charge regimes discussed in Sec.~\ref{subsection:point_resolution_initial_dist}.
 
For a \SI{10}{\femto\coulomb} beam, as shown in \hyperref[fig:point-resolution-10fC-10pC]{Fig.~\ref*{fig:point-resolution-10fC-10pC}(a)}, a beam waist is formed at $z \approx 1.25\,\mathrm{m}$ with an rms size of about \SI{0.07}{\milli\meter} in the absence of a cathode solenoid, and shifts to $z \approx 1.35\,\mathrm{m}$ with an rms size of \SI{0.97}{\milli\meter} when the cathode solenoid is applied. For this \SI{10}{\femto\coulomb}, \SI{100}{\femto\second} beam, space charge effects are relatively weak, so the point-emission microbunch can be focused effectively in both cases. Moreover, introducing the cathode solenoid does not provide further improvement in the imaging resolution (see \hyperref[fig:point-resolution-10fC-10pC]{Fig.~\ref*{fig:point-resolution-10fC-10pC}(b)}).

For a \SI{10}{\pico\coulomb} beam, \hyperref[fig:point-resolution-10fC-10pC]{Fig.~\ref*{fig:point-resolution-10fC-10pC}(c)} shows that, without the cathode solenoid, the beam reaches a waist of \SI{0.27}{\milli\meter} at $z \approx 1.5\,\mathrm{m}$. Upstream of the waist, $\sigma_{10}$ increases rapidly because of space charge effects. With an axial magnetic field of $800\,\mathrm{Gauss}$ applied at the cathode, the waist position is almost unchanged, while the rms beam size at the waist increases to \SI{1.0}{\milli\meter}. The corresponding cathode-plane resolution is better than \SI{100}{\micro\meter} (see \hyperref[fig:point-resolution-10fC-10pC]{Fig.~\ref*{fig:point-resolution-10fC-10pC}(d)}), representing an improvement of approximately one order of magnitude.

\subsection{Trajectories of point-emission macroparticles}

\begin{figure}[t]
\centering
\includegraphics[width=0.48\columnwidth]{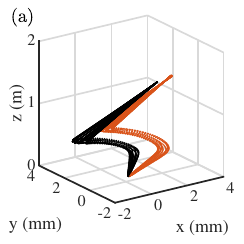}
\hfill
\includegraphics[width=0.48\columnwidth]{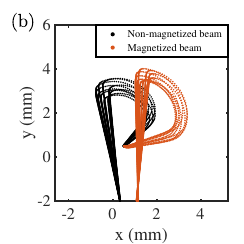}
\includegraphics[width=0.48\columnwidth]{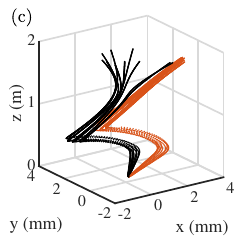}
\hfill
\includegraphics[width=0.48\columnwidth]{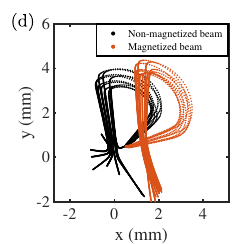}
\caption{Trajectories of 10 point-emission macroparticles. (a) 3D trajectories and (b) transverse $x$–$y$ projections for the \SI{10}{\femto\coulomb} beam. 
(c) 3D trajectories and (d) transverse $x$–$y$ projections for the \SI{10}{\pico\coulomb} beam.}
\label{fig:10fC-10pC-trajectory}
\end{figure}

The preceding analysis, based on the rms size of point-emission macroparticles and the corresponding imaging resolution, reveals the key beam dynamics features of the cathode imaging process. To further clarify how individual macroparticles are focused and subsequently diverge along the beamline, the trajectories of the point-emission macroparticles are tracked.

For the \SI{10}{\femto\coulomb} beam, \hyperref[fig:10fC-10pC-trajectory]{Figs.~\ref*{fig:10fC-10pC-trajectory}(a)} and \hyperref[fig:10fC-10pC-trajectory]{\ref*{fig:10fC-10pC-trajectory}(b)} show that the point-emission microbunch remains well focused after the imaging solenoid over the transport range from \(z=\SI{0}{\meter}\) to \(z=\SI{2}{\meter}\), for both the non-magnetized and magnetized beams.

For the \SI{10}{\pico\coulomb} beam, \hyperref[fig:10fC-10pC-trajectory]{Figs.~\ref*{fig:10fC-10pC-trajectory}(c)} and \hyperref[fig:10fC-10pC-trajectory]{\ref*{fig:10fC-10pC-trajectory}(d)} show that, in the non-magnetized imaging scheme, the trajectories of these macroparticles diverge strongly near the waist, where the rms beam size reaches its minimum and space charge effects are strongest. In contrast, when an axial magnetic field of $800\,\mathrm{Gauss}$ is applied at the cathode, the macroparticles largely preserve their initial distribution, and the transverse divergence is significantly reduced.

\subsection{Transverse momentum evolution}

\begin{figure*}[t]
\centering
\includegraphics[width=0.48\textwidth]{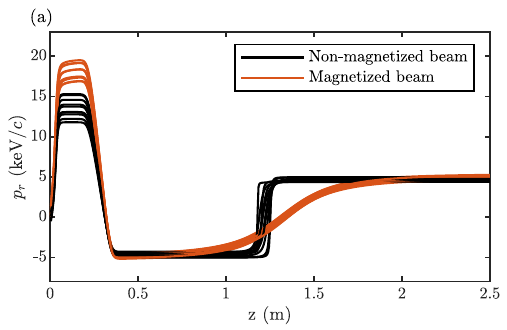}
\hfill
\includegraphics[width=0.48\textwidth]{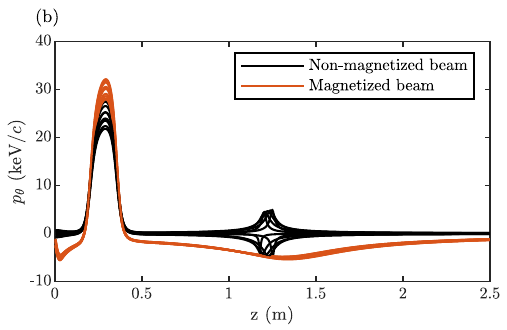}
\includegraphics[width=0.48\textwidth]{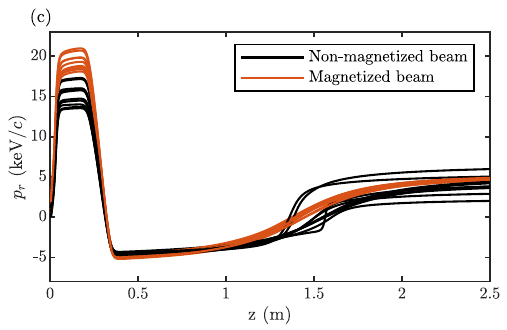}
\hfill
\includegraphics[width=0.48\textwidth]{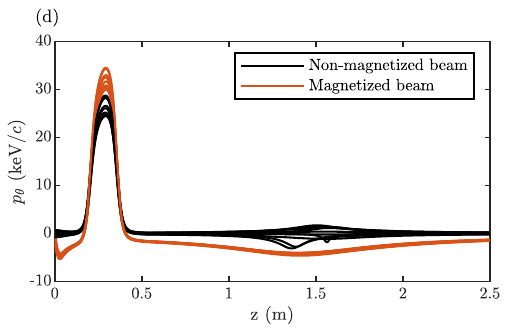}
\caption{Transverse momentum of the 10 point-emission macroparticles.
(a) Radial $p_r$ and (b) azimuthal $p_\theta$ for the \SI{10}{\femto\coulomb} beam. 
(c) Radial $p_r$ and (d) azimuthal $p_\theta$ for the \SI{10}{\pico\coulomb} beam.}
\label{fig:momentum}
\end{figure*} 

To further clarify the divergence observed in the trajectories, the evolution of the transverse momentum of the 10 macroparticles is next examined. In the cathode imaging process, several mechanisms contribute to the spread in transverse momentum. First, the electrons emitted from the cathode carry a distribution of transverse momentum due to the finite thermal emittance of the photocathode. Second, the solenoid field imposes both rotation and transverse focusing on the beam, thereby redistributing the transverse phase space. Third, space charge effects during transport further enhance this spread. As a result, even for ideal point-emission macroparticles at the cathode, the accumulated transverse momentum spread eventually produces a finite spot size on the screen and thus limits the achievable imaging resolution.

We analyze the transverse momentum in cylindrical coordinates $(r,\theta,z)$. The electron momentum is decomposed into the radial, azimuthal, and longitudinal components, $p_{r}$, $p_{\theta}$, and $p_{z}$, and the evolution of the transverse components $p_r$ and $p_\theta$ along the beamline is shown in Fig.~\ref{fig:momentum}. Assuming that the macroparticles undergo the same net rotation during imaging, integrated radial momentum spread $\int \Delta p_r\,dz$ is used as a qualitative indicator of the resolution.

The beam rotation induced by the cathode magnetic field is first examined. The cathode solenoid and the imaging solenoid are configured to provide axial magnetic fields in the same direction. The beam emitted within the cathode solenoid acquires an initial clockwise azimuthal momentum. At the entrance of the imaging solenoid, the radial magnetic field component $B_r$ reverses direction, and the azimuthal momentum is reversed accordingly, as illustrated in \hyperref[fig:momentum]{Figs.~\ref*{fig:momentum}(b)} and \hyperref[fig:momentum]{\ref*{fig:momentum}(d)}.

For the non-magnetized beam, the transverse kicks at the entrance and exit of the solenoid cancel each other, and the net azimuthal momentum vanishes in the downstream drift. This behavior is confirmed in \hyperref[fig:momentum]{Fig.~\ref*{fig:momentum}(b)}, where the azimuthal momentum of the macroparticles is essentially zero for $z > \SI{0.5}{\meter}$. When the cathode magnetic field is applied, a residual clockwise azimuthal momentum remains at the exit of the imaging solenoid. This originates from the canonical angular momentum imparted by the cathode field during emission and is preserved throughout the beamline.

Moreover, as discussed in Sec.~\ref{subsection:point_resolution}, the space charge effects near the waist play a key role in determining the imaging resolution. The momentum variation near the waist (Fig.~\ref{fig:momentum}) together with the corresponding transverse trajectories is therefore analyzed (Fig.~\ref{fig:10fC-10pC-trajectory}).

For the \SI{10}{\femto\coulomb} beam, space charge effects are negligible. As shown in \hyperref[fig:momentum]{Figs.~\ref*{fig:momentum}(a)} and \hyperref[fig:momentum]{\ref*{fig:momentum}(b)}, no pronounced distortion is observed in the transverse momentum.

For the \SI{10}{\pico\coulomb} beam, space charge effects produce a pronounced spread among the radial momentum curves. By contrast, when the cathode solenoid is applied, the radial momentum curves nearly overlap in the downstream drift. In this case, the beam waist is about \SI{1.0}{\milli\meter}, space charge effects are relatively weak, and the macroparticle trajectories show no significant transverse spread (see \hyperref[fig:momentum]{Figs.~\ref*{fig:momentum}(c)} and \hyperref[fig:momentum]{\ref*{fig:momentum}(d)}).

Without the cathode magnetic field, the radial momentum of the macroparticles changes sign from negative to positive around $z = 1.50\,\mathrm{m}$ because the macroparticles reach their focal points at slightly different axial positions. When the electrons pass through the origin, the radial unit vector $\hat{\mathbf e}_r$ reverses direction in cylindrical coordinates and the radial momentum is reversed accordingly.

\section{\label{section:QE_mapping}Beam dynamics simulation of QE mapping}

\begin{figure}[b]
\centering
\includegraphics[width=0.6\columnwidth]{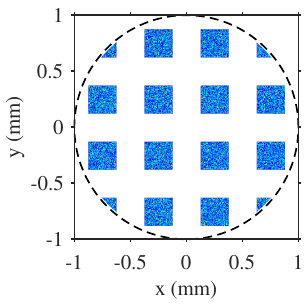}
\caption{
Initial electron beam distribution with a grid pattern in the GPT simulation.
}
\label{fig:GPT_mesh_grid}
\end{figure}

In this section, beam dynamics simulations are performed to demonstrate the cathode imaging performance for high charge beams and to investigate the effects of bunch charge and the cathode magnetic field on imaging quality. A pattern beam is generated on the cathode, as shown in Fig.~\ref{fig:GPT_mesh_grid}, with electrons uniformly distributed within each square of \SI{250}{\micro\meter} side length and a \SI{500}{\micro\meter} spacing between adjacent squares. To better simulate the cathode emission process of a beam with transverse nonuniformity, the simulations are carried out using the GPT code \cite{pulsar_gpt_code}.

\subsection{Impact of space charge effects}

\begin{figure}[tbp]

\noindent
\hbox to \columnwidth{
\includegraphics[width=0.48\columnwidth]{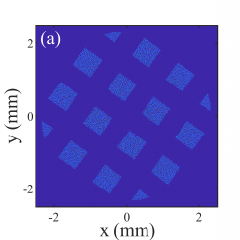}
\hfill
\includegraphics[width=0.48\columnwidth]{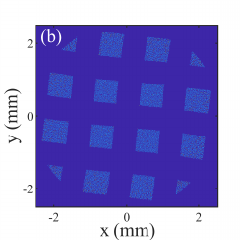}
}

\noindent
\hbox to \columnwidth{
\includegraphics[width=0.48\columnwidth]{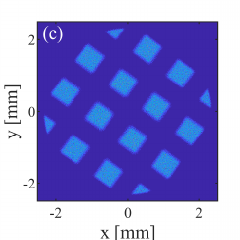}
\hfill
\includegraphics[width=0.48\columnwidth]{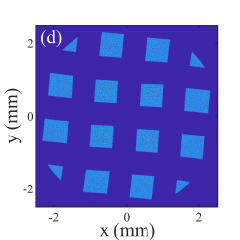}
}

\noindent
\hbox to \columnwidth{
\includegraphics[width=0.48\columnwidth]{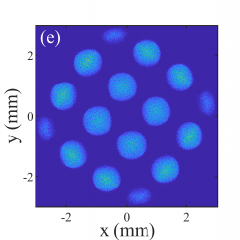}
\hfill
\includegraphics[width=0.48\columnwidth]{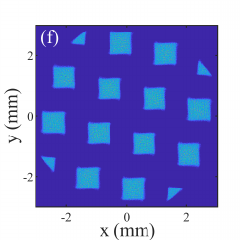}
}

\noindent
\hbox to \columnwidth{
\includegraphics[width=0.48\columnwidth]{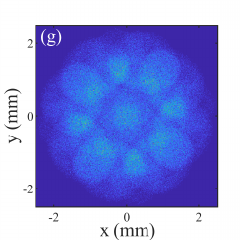}
\hfill
\includegraphics[width=0.48\columnwidth]{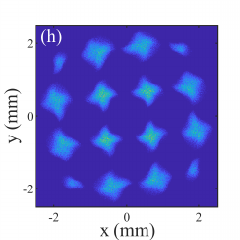}
}

\caption{
Electron distributions on the screen for a \SI{100}{\femto\second} beam with different bunch charges:
(a,b) \SI{10}{\femto\coulomb}, (c,d) \SI{100}{\femto\coulomb},
(e,f) \SI{1}{\pico\coulomb}, and (g,h) \SI{10}{\pico\coulomb}.
Panels (a,c,e,g) show the non-magnetized beam, and panels (b,d,f,h) show the magnetized beam.
}
\label{fig:space_charge_effect}
\end{figure}

In the simulations, the bunch length is fixed at $\sigma_t = \SI{100}{\femto\second}$, while the bunch charge is varied to represent different space charge regimes. Four cases are considered: \SI{10}{\femto\coulomb}, \SI{100}{\femto\coulomb}, \SI{1}{\pico\coulomb}, and \SI{10}{\pico\coulomb}. The electron distributions on the screen at the optimal imaging plane for each case are shown in Fig.~\ref{fig:space_charge_effect}.

For the \SI{10}{\femto\coulomb} and \SI{100}{\femto\coulomb} cases,
clear imaging is obtained for both the non-magnetized and magnetized beams,
with no noticeable difference in image quality. In contrast, for the \SI{1}{\pico\coulomb} and \SI{10}{\pico\coulomb} cases, clear imaging cannot be achieved with the non-magnetized beam. As discussed in Sec.~\ref{section:point_emission}, this is because the beam waist becomes too small, leading to severe distortion of the transverse distribution and preventing image formation downstream of the waist. For the magnetized beam, the increased waist size mitigates the space-charge-induced degradation of the transverse distribution, thereby improving the resolving capability of the imaging system.

In addition, for the \SI{1}{\pico\coulomb} and \SI{10}{\pico\coulomb} cases, the square bunch pattern defined at the cathode becomes slightly distorted at the imaging plane and is rotated relative to the original grid structure. This occurs under a rather extreme condition: the bunch length is only \SI{100}{\femto\second}, and the imposed grid pattern produces a highly nonuniform transverse charge density. As a result, nonlinear space charge forces can become significant and lead to the observed pattern distortion. In realistic QE mapping beams, the transverse distribution is expected to vary more gradually, so such pronounced distortion is unlikely to occur.

\subsection{Impact of cathode magnetic field}

As indicated in Sec.~\ref{section:matrix_analysis}, increasing the cathode magnetic field strength leads to an increase in the beam waist size. This change modifies the space charge evolution along the beamline: with a stronger cathode field, the enlarged waist further suppresses space charge effects, thereby potentially improving the resolving capability.

\begin{figure}[b]

\noindent
\hbox to \columnwidth{%
\includegraphics[width=0.48\columnwidth]{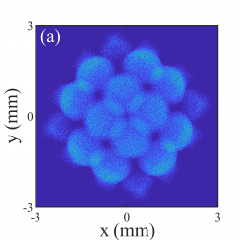}%
\hfill
\includegraphics[width=0.48\columnwidth]{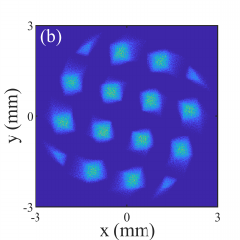}%
}

\noindent
\hbox to \columnwidth{%
\includegraphics[width=0.48\columnwidth]{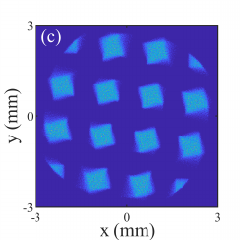}%
\hfill
\includegraphics[width=0.48\columnwidth]{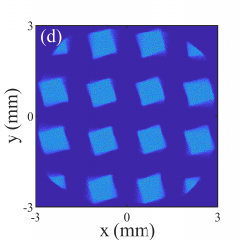}%
}

\caption{
Electron distributions on the screen under different cathode magnetic field strength:
(a) without cathode magnetic field, (b) $400\,\mathrm{Gauss}$, (c) $800\,\mathrm{Gauss}$, (d) $1200\,\mathrm{Gauss}$.}
\label{fig:cathode-solenoid-strength}
\end{figure}

In the following simulations, the on-axis peak magnetic field of the imaging solenoid is fixed to $520\,\mathrm{Gauss}$. A \SI{10}{\pico\coulomb}, \SI{3}{\pico\second} electron beam is considered, and its initial bunch distribution is shown in Fig.~\ref{fig:GPT_mesh_grid}.

For the non-magnetized beam, \hyperref[fig:cathode-solenoid-strength]{Fig.~\ref*{fig:cathode-solenoid-strength}(a)} shows that space charge effects blur the grid edges, indicating that no valid imaging condition exists in this case. The cathode magnetic field is then set to $400\,\mathrm{Gauss}$, $800\,\mathrm{Gauss}$, and $1200\,\mathrm{Gauss}$, and the corresponding results are presented in \hyperref[fig:cathode-solenoid-strength]{Figs.~\ref*{fig:cathode-solenoid-strength}(b)}-\hyperref[fig:cathode-solenoid-strength]{\ref*{fig:cathode-solenoid-strength}(d)}.

\begin{figure}[t]
\centering
\includegraphics[width=\columnwidth]{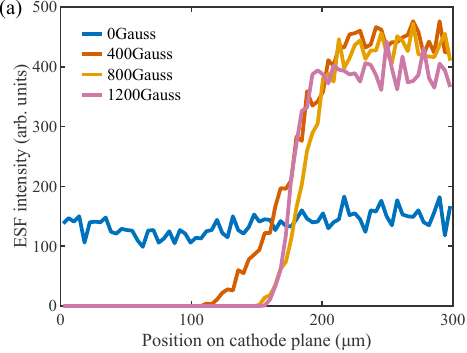}
\par
\vspace{1em}
\includegraphics[width=\columnwidth]{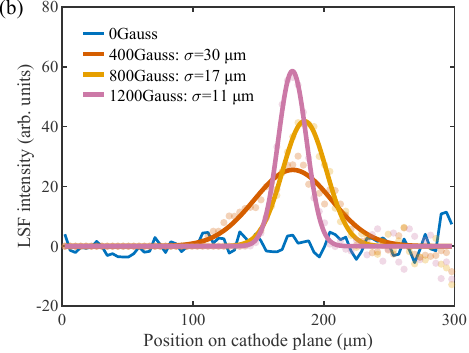}
\caption{Simulated (a) ESF and (b) LSF curves obtained at different cathode magnetic field strengths.}
\label{fig:cathode-B-vs-ESF-LSF}
\end{figure}

To evaluate the corresponding cathode-plane resolution, the electron distribution on the screen is first rotated and rescaled to the same transverse size as that on the cathode. A \SI{0.3}{\milli\meter} $\times$ \SI{0.3}{\milli\meter} rectangular region on the cathode plane is selected. The electron density is projected onto the $x$ axis with a resolution of \SI{4}{\micro\meter}/pixel. The resulting edge spread function (ESF) for each case is shown in \hyperref[fig:cathode-B-vs-ESF-LSF]{Fig.~\ref*{fig:cathode-B-vs-ESF-LSF}(a)}. Differentiation of each ESF yields the corresponding line spread function (LSF). Because the number of initial macroparticles in the GPT simulations is limited, moving average smoothing is applied to the derivative before performing a Gaussian fit. The LSFs across the grid bar edges and the corresponding Gaussian fits are shown in \hyperref[fig:cathode-B-vs-ESF-LSF]{Fig.~\ref*{fig:cathode-B-vs-ESF-LSF}(b)}. As the cathode magnetic field increases, the bar edges become progressively sharper and the imaging resolution improves. For cathode magnetic fields of \SI{400}{Gauss}, \SI{800}{Gauss}, and \SI{1200}{Gauss}, the fitted \(1\sigma\) resolutions are \SI{30}{\micro\meter}, \SI{17}{\micro\meter}, and \SI{11}{\micro\meter}, respectively.

\section{\label{section:summary}Summary}

In this paper, we propose a novel cathode imaging method based on a magnetized electron beam for photocathode QE mapping. We present matrix-based theoretical analysis and beam dynamics simulations to demonstrate its feasibility. It is shown that the magnetized beam forms a larger beam waist in the cathode imaging beamline,
which weakens space charge effects and enables point-to-point cathode imaging at high bunch charge. As the cathode magnetic field strength increases, the beam waist becomes larger, and the imaging resolution at a given bunch charge is correspondingly improved. In particular, for a \SI{10}{\pico\coulomb}, \SI{3}{\pico\second} beam with a cathode magnetic field of \SI{1200}{Gauss}, simulations indicate an imaging resolution of \SI{11}{\micro\meter}, representing nearly an order-of-magnitude improvement over the non-magnetized beam method. Therefore, magnetized beam cathode imaging offers a promising approach to rapid, high resolution, single-shot QE mapping in the high charge regime.

\begin{acknowledgments}
 This work is supported by the National Key Research and Development Program (NKRDP) of China under Grant No. 2024YFB2807501. This work is also supported by the National Natural Science Foundation of China (NSFC) under Grant No. 12275149.
\end{acknowledgments}

\bibliography{mybibfile.bib}

\end{document}